\documentclass[aps,pre,twocolumn,showpacs,superscriptaddress,groupedaddress]{revtex4}  
\usepackage{graphicx,subfigure}  
\usepackage{dcolumn}   
\usepackage{bm}        
\usepackage{amssymb}   
\usepackage{xcolor}
\usepackage[symbol]{footmisc}
\newif\ifHighlitedChanges
\def\ifHighlitedChanges{\iftrue}
\ifHighlitedChanges
  
  \def\STRIKE#1{{\color{black}\sout{#1}}}
\else
  
  \def\STRIKE#1{\relax}
\fi

\begin{document}



\title{Polymorph selection during crystallization of a model colloidal fluid with a free energy landscape containing a metastable solid}
\author{Mantu Santra}
\email{mantu.santra@gmail.com}
\affiliation{Laufer Center for Physical and Quantitative Biology, Stony Brook University, Stony Brook, New York 11794, USA}
\author{Rakesh S. Singh\footnote[3]{Current Address: Department of Chemistry, Johns Hopkins University, Baltimore, MD 21218, USA}}
\affiliation{Department of Chemical and Biological Engineering, Princeton University, Princeton, New Jersey 08544, USA}
\author{Biman Bagchi}
\email{profbiman@gmail.com}
\affiliation{Solid State and Structural Chemistry Unit, Indian Institute of Science, Bangalore 560012, India}

\date{\today}

\begin{abstract}
The free energy landscape responsible for crystallization can be complex even for relatively simple systems like hard sphere and charged stabilized colloids. In this work, using hard-core repulsive Yukawa model, which is known to show complex phase behavior consisting of fluid, FCC and BCC phases, we studied the interplay between the free energy landscape and polymorph selection during crystallization. When the stability of the BCC phase with respect to the fluid phase is gradually increased by changing the temperature and pressure at a fixed fluid-FCC stability, the final phase formed by crystallization is found to undergo a switch from the FCC to the BCC phase, even though FCC remains thermodynamically the most stable phase. We further show that the nature of local bond-orientational order parameter fluctuations in the metastable fluid phase as well as the composition of the critical cluster depend delicately on the free energy landscape, and play a decisive role in the polymorph selection during crystallization. 
\end{abstract}
\maketitle

\section {Introduction}
 The remarkable argument of Alexander and McTague, based on the Landau theory, that in the case of a simple fluid undergoing weak first order phase transition, the body-centered-cubic (BCC) phase should be favored over the thermodynamically stable face-centered-cubic (FCC) phase~\cite{mctag} has till now defied a quantitative understanding. It has been hard to reconcile with the classical nucleation theory (CNT)~\cite{becker-doring-1935, frenkel-book-1955,deben_book,bagchi_book}, which in its simplest form, considers the competition between the free energy gain in the  fluid to solid transformation and the cost of creating the fluid-solid interface; the net free energy change controls the crystallization process.  CNT may capture the kinetics for the cases where only two free energy basins (the metastable parent and the stable daughter phases) are involved, but it could  fail to address phase transformation in the presence of multiple solid phases.

The participation of the intermediate metastable phase(s) through wetting of the stable phase nuclei is an important factor that needs to be taken into account. In such cases, we have a competition between thermodynamic (the stability of solids) versus kinetic (the free energy barrier of nucleation) control. Empirically, one can invoke the Ostwald's step rule~\cite{ostwald_1} that states that the crystal phase that forms out of metastable melt need not be the thermodynamically most stable phase, rather it is the one closest in stability to the parent phase. However, the Ostwald's step rule still lacks a solid theoretical foundation. Stranski and Totomanow~\cite{ostwald_2} argued that the solid with lowest free energy barrier will emerge from the metastable fluid irrespective of its stability with respect to other solid phases, which is yet to be demonstrated in a fully microscopic computational study. 

Recent advancement in experimental techniques has led tremendous interest in understanding and controlling the phase transformation and polymorph selection in complex materials and biological systems~\cite{chung_nat_phys,delhommelle_2011,rengarajan_pccp,yoreo_science_2014, sleutel_pnas_2014, pablo_ice_2015, kratzer_softmater_2015, ritchai_natmat_2015, aizenberg_pnas_2018,sleutel_nature_2018}. Complex materials are often characterized by the presence of multiple length and energy scales in the interaction potential between the constituents, and usually display complex energy landscape~\cite{wales_book} and 
rich phase behavior consisting of multiple phases~\cite{mat_1,mat_2,mat_3,lekkerkerker_2002,bagchi_2dm,bagchi_ss_2013,pablo_nature_2014,sciortino_natphys_2014}. Understanding the nature of the free energy landscape 
and its connection with the pathways of phase transition has great practical 
relevance in seemingly distinct branches of science ranging from 
materials (e.g. polymorph selection~\cite{chung_nat_phys,bagchi_osr_2013}) to biological  
(e.g. protein crystallization, aggregation~\cite{frenkel_science,stradner_nature_2004,sleutel_pnas_2014}) sciences. Crystal engineering relies heavily on gaining a molecular level understanding and control of the free energy landscape, and in turn, the pathways of phase transformation~\cite{sleutel_pnas_2014,sleutel_nature_2018}.  Multiple intermediate metastable phases are already known to play an important role in the formation of random spin and structural glasses in both experimental and theoretical descriptions~\cite{villain, ted_wolynes,xia_wolynes,harrowell_2003}.

In experiments, proper characterization and controlled change of the complex free energy landscape pose a major challenge in gaining fundamental understanding of the crystallization and polymorph selection processes. For example, any change in thermodynamic conditions such as temperature ($T$) and pressure ($P$) alters the whole free energy landscape (the stability of all the phases), and thus prevents us from understanding how the relative stability of a particular intermediate metastable polymorph would affect the pathways and rate of crystallization. Using phenomenological classical density functional theory (DFT)~\cite{bagchi_book}, we recently showed that the controlled changes of the (meta)stability of intermediate phases can give rise to diverse non-classical pathways of phase transformation, ranging from wetting-mediated to Ostwald's step rule-like scenario~\cite{bagchi_osr_2013,bagchi_ice_2014}. However, precise microscopic pathways and criteria for polymorph selection in such complex systems largely remain elusive and demand further controlled studies with atomic resolutions.

In this work, we used hard-core repulsive Yukawa as a model system to computationally explore the polymorph selection during crystallization. This system exhibits a rich phase behavior consisting of fluid-FCC/BCC and FCC-BCC phase coexistence lines along with two triple points~\cite{meijer_1997}, and thus, ideally suited to study the interplay between the free energy landscape and the selection of FCC and BCC polymorphs. We explored the diverse pathways of phase transformation through the controlled change of the free energy landscape. On gradually increasing the stability of the BCC phase with respect to the fluid phase at a fixed fluid-FCC stability, we observed a cross-over from the formation of the thermodynamically most stable FCC phase through a wetting-mediated pathway to an Ostwald's step rule-like scenario where the BCC phase of intermediate stability (stable with respect to the fluid and metastable with respect to the FCC) grows despite FCC being the thermodynamically the most stable phase. We further observed that the composition of the critical cluster depends delicately on the free energy landscape, and plays a decisive role in the polymorph selection during crystallization. Additionally, we also explored the microscopic pathways of the emergence of the composition of different (FCC and BCC) polymorphs in the critical cluster from the metastable fluid phase.

\section {Model and Method Details} \label{method}
\subsection{Model details}
We performed Monte-Carlo (MC) simulations~\cite{frenkel_book} on a system interacting via hard-core repulsive Yukawa potential, 
\begin{equation}\label{eq:potential}
  \beta U(r) = \left\{
    \begin{array}{lc}
      \infty, &  r \le \sigma  \\
       \beta \epsilon \frac{\exp \left[-k(r/\sigma - 1)\right]}{r/\sigma}, &   ~~r > \sigma {\rm \;\; }
     \end{array}
   \right.\;,
\end{equation}                                       		
where $\sigma$ is the particle diameter and $\epsilon$ is the energy at contact distance $\sigma$. $\beta= 1/k_{\rm B}T$ where $k_{\rm B}$ is the Boltzmann constant and $T$ is the temperature. $\sigma$ and $\epsilon$ are used as units of length and energy, respectively. In this study, we have truncated 
the interaction potential at a distance $r_c = 3.0\sigma$ and shifted to zero, and chose $k\sigma=5$ as this value is commonly used in majority of the studies on Yukawa system~\cite{meijer_1997, kratzer_softmater_2015}, and the phase diagram for this parameter choice suits best to this work (note that, the phase behavior of Yukawa system shows strong dependence on the choice of the potential parameters~\cite{dijstra_2003}).    

\subsection{Computation of phase diagram}\label{subsec:phase}
To obtain the phase diagram, we first computed the Helmholtz free energy ($F$) of FCC and BCC phases at $\beta=8$ and $\rho\sigma^3=0.75$ using thermodynamic integration~\cite{frenkel_book} in \textit{NVT} ensemble for a system consisting of $N = 432$ particles for BCC and $N = 500$ particles for FCC. Using these Helmholtz free energies, we obtained the chemical potential at different pressures by integrating the equation of state~\cite{frenkel_book}. The chemical potential of the fluid phase was computed using Widom insertion method~\cite{frenkel_book} at $\beta=8$ and $\rho \sigma^3=0.15$. Again, integration over the equation of state was performed to determine the chemical potential of the fluid phase as a function of pressure. Equating these chemical potentials of fluid and solid (BCC and FCC) phases we obtained the coexistence pressures $\beta P \sigma^3= 26.5$ and $27.4$ at $\beta=8$ for fluid-BCC and fluid-FCC, respectively. We verified these coexistence pressures using direct free energy calculation employing umbrella sampling~\cite{Verrocchio_2012} and obtained $\beta P \sigma^3=26.8$ and $27.7$ for fluid-BCC and fluid-FCC, respectively. After computing the coexistence fluid-solid pressures at $\beta=8$, we obtained the fluid-BCC and fluid-FCC coexistence pressures as a function of $\beta$ using the Clausius-Clapeyron relation~\cite{kofke_jcp_1993,kofke_molphys_1993},
\begin{equation}\label{eq:CC}
P_2 = P_1\exp\left[\frac{(\beta_2 - \beta_1)\Delta h}{\beta_1 P_1 \Delta v}\right], 
\end{equation}
where $\Delta h = h_j(\beta_1, P_1) - h_i(\beta_1, P_1)$ is the enthalpy difference per particle between phases $j$ and $i$, and $\Delta v = v_j(\beta_1, P_1) - v_i(\beta_1, P_1)$ is their volume difference per particle at inverse temperature $\beta_1$ and pressure $P_1$. $P_1$ is the coexistence pressure at inverse temperature $\beta_1$, and $P_2$ is the coexistence pressure at inverse temperature $\beta_2$. We carried out \textit{NPT} MC simulations with $20,000$ equilibrium MC steps (1 MC step equals $N$ numbers of single particle displacement and one volume move attempts) followed by $50,000$ production steps at $\beta_1$ and $P_1$ in both the phases, $i$ and $j$, and computed $\Delta h$ and $\Delta v$. The systems contained $432$ particles for fluid-BCC and $500$ for fluid-FCC coexistence lines. Using Eq.~\ref{eq:CC}, we obtained fluid-BCC and fluid-FCC coexistence pressures as a function of $\beta$ and they cross each other at a triple point, $\beta=4.5$ and $\beta P \sigma^3= 26.2$. Starting from this triple point we computed the BCC-FCC coexistence line as a function of $\beta$ using Clausius-Clapeyron equation as described above. Finally, combining these three coexistence lines --- fluid-BCC, fluid-FCC and BCC-FCC --- we obtained the phase diagram shown in Fig.~\ref{fig_1}. The chemical potential differences of FCC and BCC phases with respect to the fluid phase at thermodynamic conditions studied in this work (asterisks in Fig.~\ref{fig_1}, and Table~\ref{tab:table1}) were computed using thermodynamic integration for larger systems consisting of $N=2662$ and $2916$ particles for BCC and FCC phases, respectively. 

\subsection{Identification of solid-like particles and polymorphs}\label{subsec:polymorph}
Solid-like crystallites in the metastable fluid phase were identified using the method introduced by Frenkel and co-workers~\cite{frenkel_1996}. This method first identifies the local bond-orientational symmetry of particle $i$ using a complex vector $q_{lm}(i)$~\cite{bop} as ,
\begin{equation}\label{eq:1}
 q_{lm}(i) = \frac{1}{N_b(i)}\sum_{j=1}^{N_b(i)}Y_{lm}(\mathbf{r_{ij}})
\end{equation}
where $N_b(i)$ is the number of nearest neighbors of the $i^{th}$ particle. Two particles were considered to be neighbors if the distance between them ($|\mathbf{r_{ij}}|$) was less than the cut-off distance of $q_c=1.38/(\rho \sigma^3)^{1/3}$, where $q_c$ is the radius of the first shell of FCC lattice (measured from the position of the minimum separating the first and second peaks in the radial distribution function), $\rho \sigma^3$ is the reduced density. $Y_{lm}(\mathbf{r_{ij}})$ is the spherical harmonics and $\mathbf{r_{ij}}$ is the distance vector between the particle $i$ and its neighbor $j$. $l$ and $m$ are integers with $-l \le m \le l$. The unit vector of $q_{lm}(i)$ is given by,
\begin{equation}
d_{lm}(i) = \frac{q_{lm}(i)}{\Big(\sum_{m=-l}^{m=l} |q_{lm}(i)| ^2\Big)^{1/2}}.
\end{equation}
Using the unit vector $d_{lm}(i)$, a scalar product $S_l(i,j)$ which measures the correlation in bond orientational order between neighboring particles can be defined as,
\begin{equation}
S_l(i,j) = \sum_{m=-l}^{m=l} d_{lm}(i).d_{lm}^*(j),  
\end{equation}
where the superscript $^*$ indicates complex conjugate. Two neighboring particles $i$ and $j$ are considered to be connected if $S_6(i,j) > 0.7$. The particle $i$ is identified as solid-like if the number of such connections is more than $7$. 

In order to assign the polymorphic identity of a solid-like particle, we employed the coarse-grained (or locally averaged) bond orientational order parameter introduced by Lechner and Dellago~\cite{lechner_jcp_2008}. Using the order parameter given in Eq.~\ref{eq:1}, one can define a locally averaged complex vector $\bar{q}_{lm}(i)$ as,
\begin{equation}\label{eq:2}
 \bar{q}_{lm}(i) = \frac{1}{N_b(i) + 1}\sum_{j=0}^{N_b(i)} q_{lm}(j)
\end{equation}
where $j=0$ indicates the particle $i$ itself. Given the coarse-grained complex vector $\bar{q}_{lm}(i)$, one can further define coarse-grained order parameters  $\bar{q}_{l}(i)$ and $\bar{w}_{l}(i)$ as
\begin{equation}
 \bar{q}_l(i) = \sqrt{\frac{4\pi}{2l+1}\sum_{m=-l}^{l}|\bar{q}_{lm}(i)|^2}
\end{equation}
and 
\begin{equation}
\bar{w}_l(i) = \sum_{m_1 + m_2 + m_3 = 0} \left(\begin{array}{clcr}
l & l & l \\
m_1    & m_2 & m_3                                        
\end{array}\right) \frac{\bar{q}_{lm_1}(i)\bar{q}_{lm_2}(i)\bar{q}_{lm_3}(i)}{\left[\sum_{m=-l}^{l}|\bar{q}_{lm}(i)|^2\right]^{3/2}}
\end{equation}
where the term in the parentheses $\left( ... \right)$ indicates the Wigner $3j$ symbol. The integers $m_1$, $m_2$ and $m_3$ range from $-l$ to $l$ and only the terms with $m_1+m_2+m_3=0$ are allowed to contribute to the summation. Once $\bar{w}_l(i)$ is defined, we identify a previously assigned solid-like particle as BCC-like if $\bar{w}_6 > 0$, whereas it is considered to be HCP-like if $\bar{w}_6 \le 0$ and $\bar{w}_4 > 0$. A particle is considered as FCC-like if $\bar{w}_6 \le 0$ and $\bar{w}_4 \le 0$~\cite{tanaka_2012}.

\subsection{Computation of nucleation free energy}\label{subsec:free}
The fluid to solid nucleation free energy profiles shown in Fig.~\ref{fig_2} were computed by employing umbrella sampling method~\cite{umbrella} with size of the largest cluster as the order parameter in \textit{NPT} ensemble consisting of $N = 2916$ particles. The force constant of the umbrella potential was taken to be $\lambda=0.1 k_{\rm B}T$. The fluid to FCC nucleation free energy profile was computed by biasing the system along the size of the largest FCC cluster with force constant $\lambda=0.1 k_{\rm B}T$, while simultaneously preventing the formation of BCC clusters using another umbrella potential with $\lambda=100 k_{\rm B}T$ along the size of the largest BCC-like cluster having minimum at cluster size $n_b = 0$. Similarly, for fluid to BCC nucleation free energy profile, two harmonic biasing potentials --- one along the size of the largest BCC cluster with $\lambda=0.1 k_{\rm B}T$  and the other with $\lambda=100 k_{\rm B}T$ along the size of the largest FCC cluster having minimum at cluster size $n_f = 0$ --- were employed simultaneously.

\section{Results and Discussion}\label{results}
 \subsection{Phase diagram}
\begin{figure}[!ht]
\includegraphics[width=\linewidth]{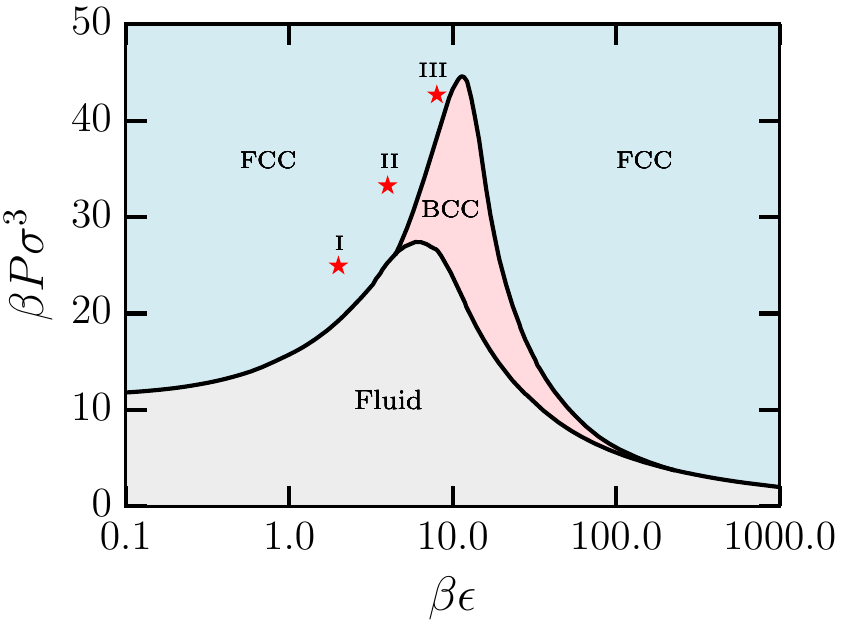}
\caption{Computed phase diagram of $k\sigma=5$ hard-core repulsive Yukawa system. The phase diagram consists of BCC-fluid, FCC-fluid and FCC-BCC coexistence lines along with two triple points. The red asterisks (marked with $\rm I$, $\rm II$ and $\rm III$) indicate the representative state points where we have performed simulations to explore the pathways of crystallization from the metastable fluid. As reported in Table~\ref{tab:table1}, at these state points, the chemical potential difference between the fluid and the FCC solid ($\beta \Delta \mu_{\rm FCC}$) is fixed at $-0.28$ and the stability of BCC phase with respect to the fluid phase ($|\beta \Delta \mu_{\rm BCC}|$) gradually increases on moving from $\rm I$ to $\rm III$.}
\label{fig_1}
\end{figure}

 In Fig.~\ref{fig_1}, we show the computed phase diagram of the $k\sigma=5$ hard-core repulsive Yukawa system (for details of the phase diagram computation see Section~\ref{subsec:phase}). The red asterisks in Fig.~\ref{fig_1} denote the thermodynamic conditions at which crystallization has been studied in this work. At these conditions, the fluid is metastable with respect to both BCC and FCC phases. FCC phase is the thermodynamically most stable 
and the BCC phase is stable with respect to the fluid phase, but metastable with respect to the FCC phase (Table~\ref{tab:table1}). As one moves from I to III via II in the phase diagram (Fig.~\ref{fig_1}), the free energy difference between the fluid and FCC remains same, however, BCC gradually becomes more and more stable with respect to the fluid phase (Table~\ref{tab:table1}). 

\begin{table}[!htbp]
\caption{\label{tab:table1} The inverse temperature ($\beta = 1/k_{\rm B}T$), reduced pressure ($\beta P \sigma^3$), reduced chemical potential difference between the fluid and the BCC solid ($\beta \Delta \mu_{\rm BCC}$), and the fluid and the FCC solid ($\beta \Delta \mu_{\rm FCC}$) at state points I, II and III in the phase diagram (Fig.~\ref{fig_1}) are reported.}
\begin{ruledtabular}
\begin{tabular}{lcrcr}
& $\beta$ & $\beta P \sigma^3$ & $\beta \Delta \mu_{\rm BCC}$ & $\beta \Delta \mu_{\rm FCC}$  \\
\hline
I & 2 & 24.9 & -0.15 & -0.28 \\
II & 4 & 33.6 & -0.19 & -0.28 \\ 
III & 8 & 42.7 & -0.26 & -0.28 \\ 
\end{tabular}
\end{ruledtabular}
\end{table}

\subsection{Nucleation free energy barrier and polymorph selection}
\begin{figure}[!htbp]
\includegraphics[width=\linewidth]{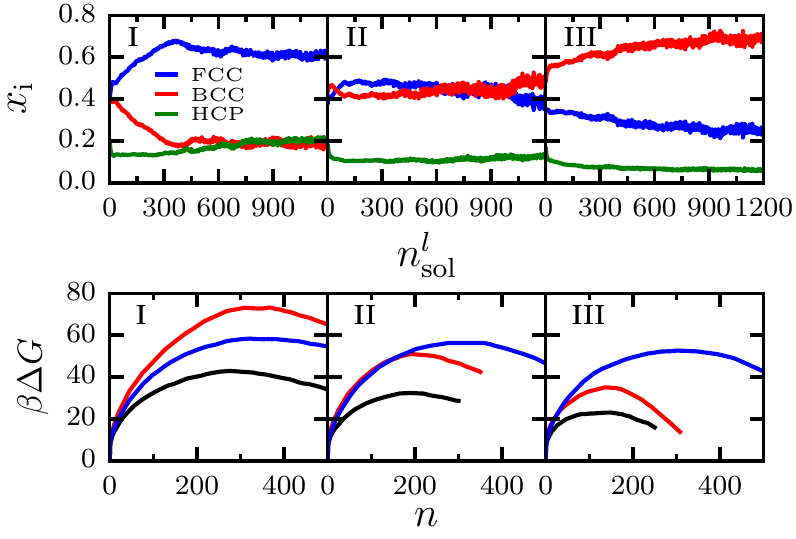}
\caption{(top) The variation of the fraction of FCC (blue), BCC (red) and HCP (green)-like particles ($x_{\rm i}$) with the total number of particles in the largest solid-like cluster ($n_{\rm sol}^l$) at state points I, II and III in the phase diagram. $x_i$ is defined as $n_{\rm i} / n_{\rm sol}^l$, where $n_{\rm i}$ is the number of particles of the $i^{th}$ polymorph in the largest cluster. Note the crossover from the FCC dominated to the BCC-dominated largest cluster. (bottom) Nucleation free energy profiles: free energy cost ($\beta \Delta G$) for the formation of FCC, BCC and solid-like clusters (indicated by blue, red and black colored lines, respectively) of size $n$ is shown. Note the crossover of the nucleation free energy barriers of FCC and BCC phases on moving from I to III.}
\label{fig_2}
\end{figure}

To uncover the role of the presence of metastable BCC phase of varying stability on the microscopic mechanism of crystallization, in Fig.~\ref{fig_2} (top panel), we show the dependence of the composition of the largest solid-like cluster (fraction of FCC, BCC and HCP-like particles in the largest solid-like cluster) on its size. We follow the method introduced by Frenkel and coworkers~\cite{frenkel_faraday} to define solid-like particles based on local bond-orientational order parameter $q_6$~\cite{bop}. The solid-like particles in the largest cluster are further identified as BCC, FCC and HCP-like based on their averaged local $\bar{w}_6$ and $\bar{w}_4$ order parameters~\cite{lechner_jcp_2008, tanaka_2012} (see Section~\ref{subsec:polymorph} for the details). A solid-like particle with $\bar{w}_6>0$ is considered as BCC-like, whereas a particle with $\bar{w}_6 \le 0$ and $\bar{w}_4>0$ as HCP-like. FCC-like particles are those with $\bar{w}_6 \le 0$ and $\bar{w}_4 \le 0$. This useful assignment criteria to distinguish different solid-like local environments in the metastable fluid phase has been used extensively in computer simulations of model atomic and molecular systems~\cite{tanaka_pnas,tanaka_gcm_2012,tanaka_natmat,xu_pre_2016}. 

At the thermodynamic conditions studied here, the metastable fluid does not undergo spontaneous phase transition on the simulation time scale. Therefore, we employed metadynamics simulations~\cite{metadynamics} considering the size of the largest solid-like cluster as order parameter to assist the system to overcome the nucleation free energy barrier and grow spontaneously~\cite{desgranges_jcp_2007,mithen_jcp_2015} (note that, as the bias is on the size of the largest solid-like cluster, it does not affect the natural selection of polymorphs). When BCC is only marginally stable with respect to the fluid phase (at I), we observe nucleation of the FCC-like clusters. On increasing the stability of the BCC phase (at II), we observe a competitive growth of the both, FCC and BCC-like clusters. On further increasing the stability of the BCC phase (at III), we observe an Ostwald's step rule like scenario where BCC phase of intermediate stability nucleates from the fluid, despite FCC being the thermodynamically most stable phase (Fig.~\ref{fig_2}, top panel). This observed  crossover from the FCC-dominated to the BCC-dominated cluster on gradual increase of the stability of the BCC phase is consistent with the predictions of our recent classical DFT~\cite{bagchi_osr_2013}. As the extent of HCP-like particles at all the three conditions --- I, II and III --- is low, from now onwards FCC denotes FCC+HCP unless HCP is explicitly specified.  

\begin{figure}[ht]
\includegraphics[width=\linewidth]{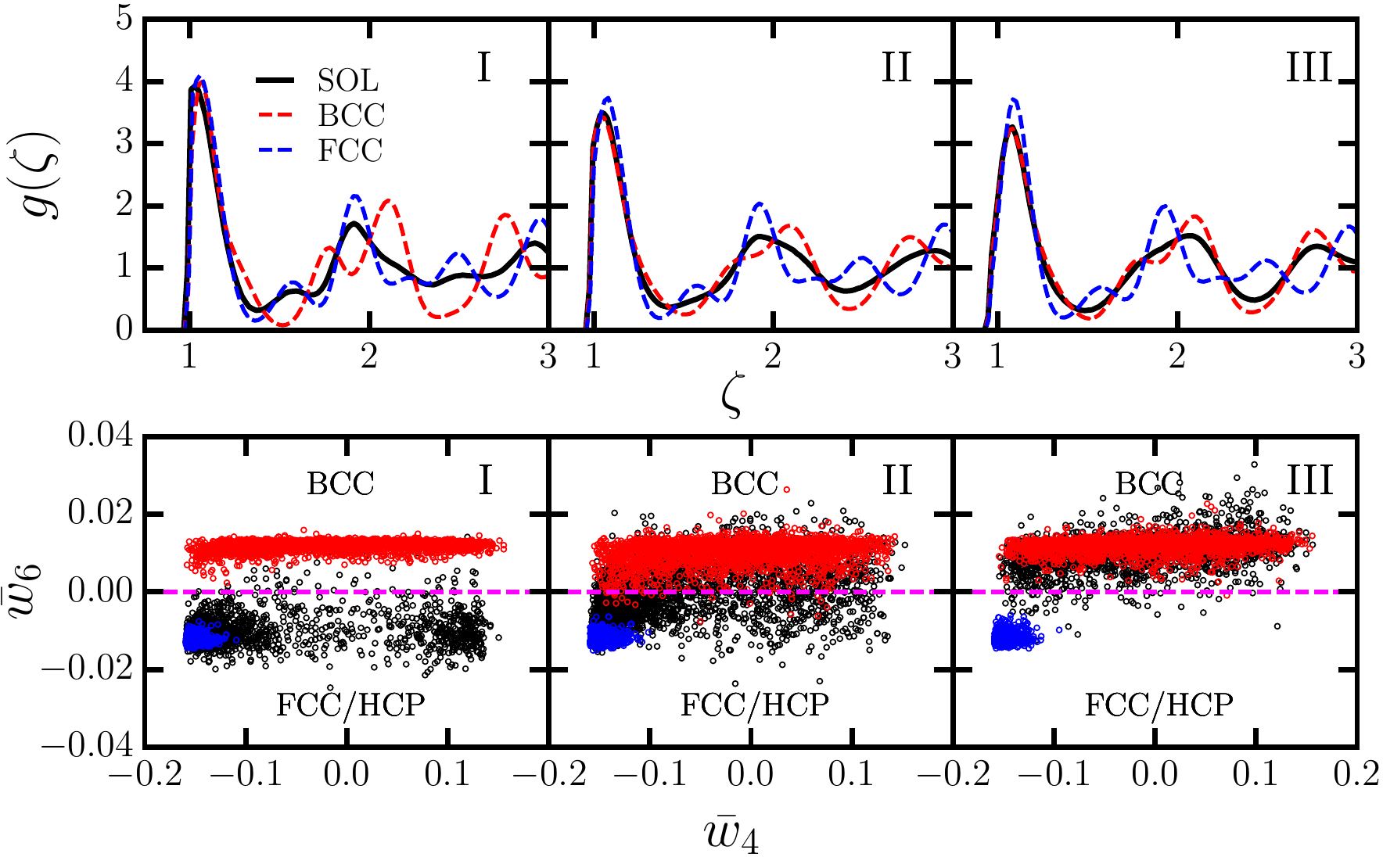}
\caption{(top) The radial distribution functions (RDFs) of the final solids formed after crystallization from the metastable fluid phase (black lines) along with the RDFs of the equilibrated pure FCC and BCC phases (indicated by dashed blue and red lines, respectively) at thermodynamic conditions of I, II and III. $\zeta = r \rho^{1/3}$ is the scaled distance. Note that, at I and III, the RDFs are similar to that of the pure thermally equilibrated FCC and BCC phases, respectively. (bottom) Scatter plot of the final solid phase formed from the metastable fluid (black circles) in the $\bar{w}_4-\bar{w}_6$ plane at I, II and III. The red and blue circles denote the BCC and FCC phases, respectively. The horizontal dashed magenta lines at $\bar{w}_6 = 0$ separate BCC-like particles from FCC/HCP-like particles. Note the formation of FCC/HCP ($\bar{w}_6 < 0$), FCC/HCP-BCC mixture and BCC ($\bar{w}_6 > 0$) solids at I, II and III, respectively.}
\label{fig_1s}
\end{figure}

Furthermore, by using umbrella sampling~\cite{umbrella, frenkel_book}, we also computed the nucleation free energy barrier for the solid-like clusters along with the FCC and BCC-like clusters (Fig.~\ref{fig_2}, bottom panel). For the FCC nucleation free energy profile, we first identified FCC and BCC-like particles using the above mentioned criteria and then suppressed BCC-like fluctuations by imposing an umbrella bias along the number of BCC-like particles ($n_{b}$) with its minimum at $n_{b}=0$. For the BCC nucleation free energy profile, the same approach was followed except that the FCC-like fluctuations were suppressed, in place of BCC (see Section~\ref{subsec:free} for the details). This approach provides conditional free energy barrier of nucleation for the pure FCC and BCC phases where the (indirect) participation of the other phases --- BCC and FCC, respectively --- is either absent or negligibly small. This conditional free energy barrier can provide an estimate of the change in the free energy barrier due to compositional heterogeneity of the critical cluster. 

As evident from Fig.~\ref{fig_2}, the free energy barrier of crystallization decreases on moving from I to III, even though the stability of FCC ($\beta \Delta \mu_{\rm FCC}$) remains the same. The fluid-to-FCC nucleation barrier (in absence of wetting by BCC-like particles) remains approximately the same --- consistent with the fixed $\beta \Delta \mu_{\rm FCC}$ at all the three conditions (Table~\ref{tab:table1}). Due to the gradual increase of the stability of the BCC phase on moving from I to III, the fluid-to-BCC nucleation free energy barrier deceases and crosses the fluid-to-FCC nucleation barrier near the thermodynamic condition of II. This crossover in the nucleation barrier leads to a transition in the nucleation mechanism from the formation of thermodynamically favored FCC phase to the formation of the metastable BCC phase (Fig.~\ref{fig_2}, top panel), unambiguously suggesting an Ostwald's step rule like scenario. 

The polymorphic identity of the final solid formed after crystallization from the metastable fluid phase at state points I, II and III (Fig.~\ref{fig_2}) was confirmed by comparing the radial distribution functions (RDFs) and the local bond-orientational order parameters ($\bar{w}_4$ and $\bar{w}_6$) of the solid phases with respective pure FCC and BCC phases equilibrated at the same thermodynamic conditions (shown in the top and bottom panels of Fig.~\ref{fig_1s}, respectively). At I and III, the RDFs and the local $\bar{w}_4$ and $\bar{w}_6$ values suggest that the final solid phases formed from the metastable fluid (indicated by the black lines in the top panel and black circles in the bottom panel) closely resemble the structures of the thermally equilibrated FCC (dashed blue line in the top panel and blue circles in the bottom panel) and BCC (dashed red line in the top panel and red circles in the bottom panel) phases, respectively. At II, however, the final solid phase can be considered as a mixture of FCC and BCC phases, which is a consequence of competitive nucleation and growth at this thermodynamic condition (Fig.~\ref{fig_2}).

\begin{figure}[!ht]
\includegraphics[width=\linewidth]{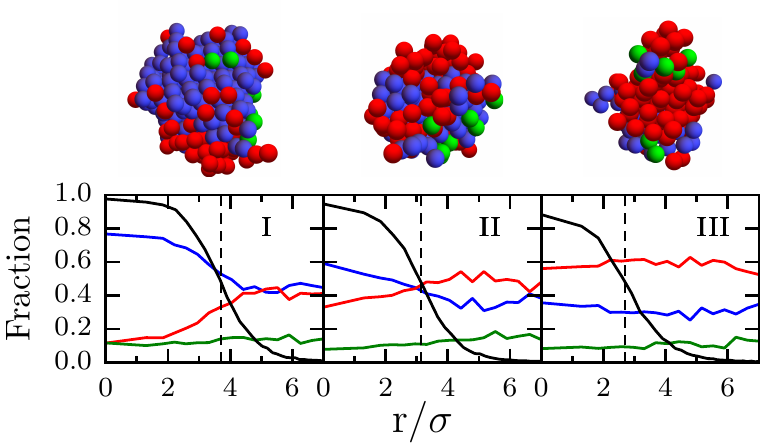}
\caption{(top) The composition of the critical cluster at state points I, II and III in the phase diagram. The blue, red and green spheres denote the FCC, BCC and HCP-like particles, respectively, in the cluster. (bottom) Composition profiles (number of particles of $i$th solid phase in a shell of radii $r$ and $r+\Delta r$ divided by the total number of solid-like particles in that shell) for FCC(blue), BCC(red) and HCP(green)-like particles in the critical cluster as a function of the distance from the center of the cluster. Solid black lines indicate the variation of the solid-like particles as a function of distance from the center of the cluster at I, II and III, respectively, and the positions of the vertical dashed black lines are the radii of the respective clusters. Fraction of solid-like particles at a distance $r$ from the center of the cluster is computed by taking the ratio of the number of solid-like particle and the total (both solid and fluid) particles in a shell of radii $r$ and $r+dr$.}
\label{fig_3}
\end{figure}

\subsection{Composition of critical clusters}
To further uncover how the composition of the critical cluster depends on the free energy landscape or the relative stability of different solid phases, in Fig.~\ref{fig_3}, we show representative snapshots along with composition profiles of the critical cluster averaged over hundred independent simulations at I, II and III. As the snapshots show, at I, the critical cluster is composed of mostly FCC-like particles with random patches of BCC-like particles at the surface only partly covering the nucleating and growing cluster. At II, the critical cluster is a mixture of both FCC and BCC-like particles of comparable fraction, and at III, we observe BCC-dominated cluster along with dispersed FCC and  HCP-like particles. 

At the bottom panel of Fig.~\ref{fig_3}, we show the composition profiles, defined as the number of particles of $i$th solid phase in a shell of radii $r$ and $r+\Delta r$ divided by the total number of solid-like particles in that shell, for FCC, BCC and HCP-like particles in the critical clusters along with the normalized density of the solid-like particles (black line) as a function of the distance from the center of the cluster. At I, the fraction of BCC-like particles increases on moving outward from the center to the surface (denoted by the vertical dashed line) of the cluster, indicating preferential wetting of the surface by the BCC phase. At II, the FCC and BCC-like particles are randomly distributed throughout the cluster in a similar proportion. At III, on the other hand, BCC-like particles dominate throughout the cluster and we do not observe any signature of wetting of the surface by the FCC phase. The lower free energy barrier of crystallization compared to the free energy barriers of pure phases (Fig.~\ref{fig_2}, bottom panel) can be attributed to the compositional heterogeneity and wetting of the solid-like clusters.  

Although recent computer simulation studies on atomic systems (such as Lennard-Jones, Gaussian core model, hard-core Yukawa and hard sphere)\cite{frenkel_prl_1995,desgranges_prl_2007,tanaka_pnas,tanaka_2012,tanaka_gcm_2012,axel_2015,xu_pre_2016,xu_2018} demonstrate various pathways of crystallization at different thermodynamic conditions, an exact criterion for the change in  mechanism of crystallization from wetting-mediated to Ostwald's step rule was still missing. In the present study, controlled change of the free energy surface (or more precisely, on changing the stability of the BCC phase at a fixed stability of the FCC phase with respect to the fluid) and the explicit computation of the conditional free energy of the pure polymorphs (Fig. 2) enabled us to gain quantitative understanding of this criterion.  

\subsection{Fluctuations of local bond-orientational order parameter in the metastable fluid}
\begin{figure}[!htbp]
\includegraphics[width=\linewidth]{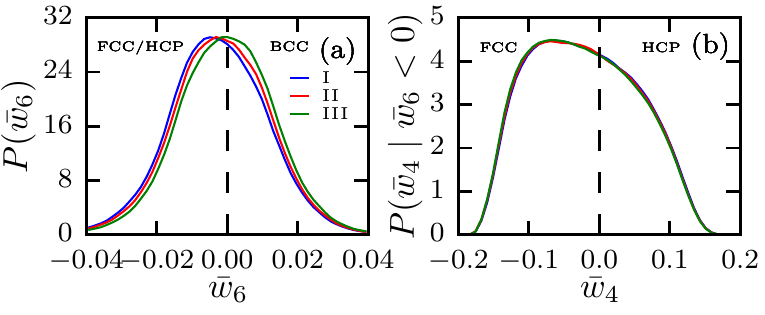}
\caption{(a) The probability distribution of the local $\bar{w}_6$ order parameter in the metastable fluid phase at I, II and III for particles with $\bar{q}_6 > 0.27$. On moving from I to III, note the gradual disappearance of the peak at negative $\bar{w}_6$ suggesting suppression of FCC-like fluctuations. (b) The conditional probability distribution of local $\bar{w}_4$ order parameter for particles with $\bar{w}_6 < 0$ and $\bar{q}_6 > 0.27$ in metastable fluid is shown. The asymmetry in the distributions suggests that the fluctuations are dominated by FCC-like ($\bar{w}_4 \le 0$) local structures.}
\label{fig_4}
\end{figure}

In a metastable fluid, through thermal fluctuations, crystallites of relatively stable phase(s) appear and disappear and sometimes grow leading to the phase transition. Recent studies show that the key to polymorph selection is hidden in the bond-orientational order parameter fluctuations in the metastable fluid~\cite{tanaka_2012,tanaka_gcm_2012,molinero_2012}. To explore how the stability of the intermediate BCC phase alters the local structural fluctuations in the metastable fluid phase, in Fig.~\ref{fig_4}, we show the computed local $\bar{w}_6$ distribution ($P(\bar{w}_6)$) along with $\bar{w}_4$ distribution of the particles with $\bar{w}_6 \le 0$ at metastable state conditions I, II and III. In computation of $P(\bar{w}_6)$ and $P(\bar{w}_4)$, we consider particles with $\bar{q}_6>0.27$~\cite{tanaka_2012}. The former enables us to distinguish BCC and HCP/FCC-like, and the latter distinguishes HCP and FCC-like local structural fluctuations. On gradually increasing the stability of the BCC phase (I$\rightarrow$III), we observe suppression of FCC/HCP-like fluctuations and concurrent enhancement of BCC-like fluctuations (Fig.~\ref{fig_4}).  

\begin{figure}[ht]
\includegraphics[width=\linewidth]{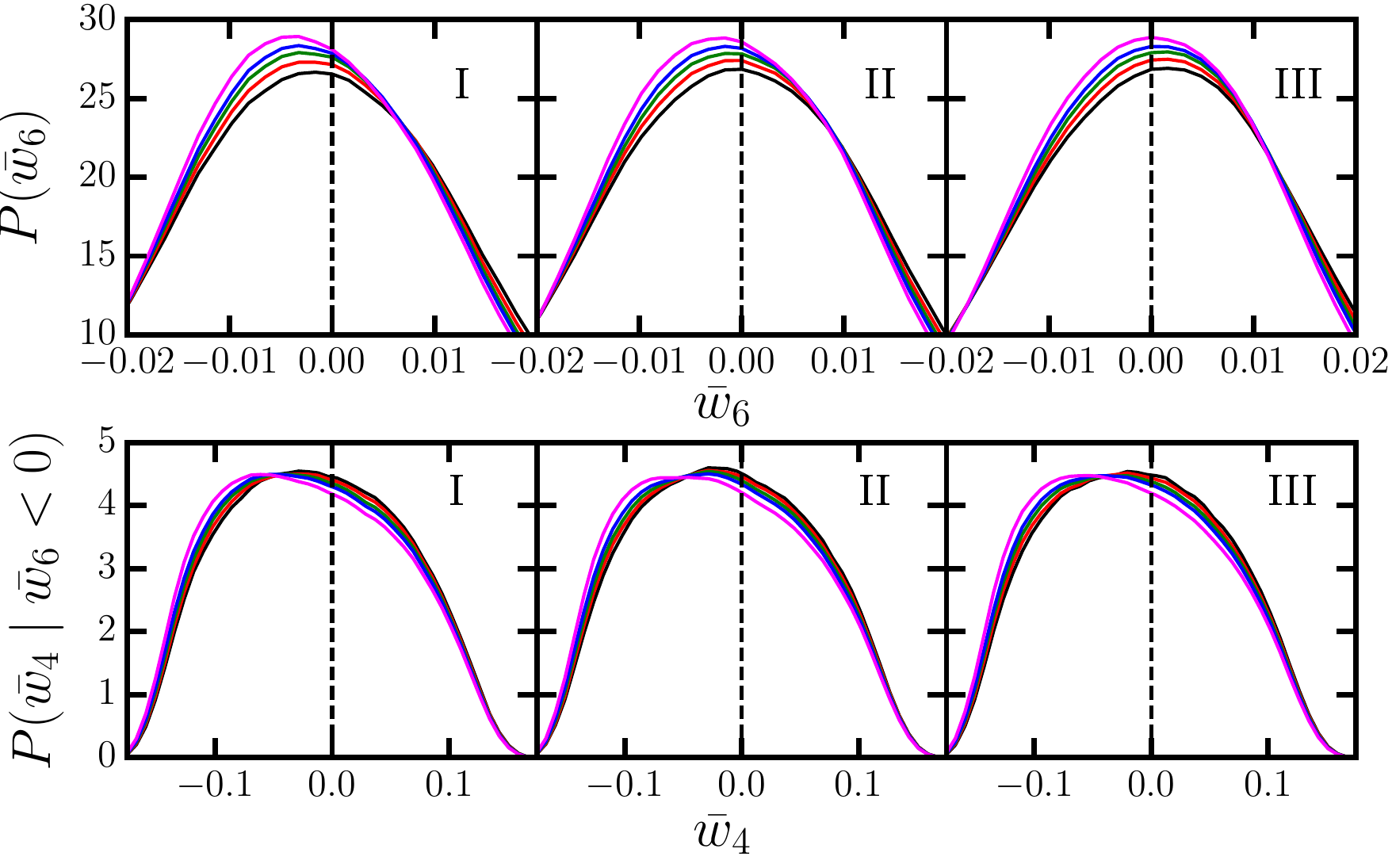}
\caption{(top) The probability distribution of order parameter $\bar{w}_6$ for the particles in the fluid phase with $0.27<\bar{q}_6<0.28$ (black), $0.27<\bar{q}_6<0.29$ (red), $0.27<\bar{q}_6<0.30$ (green), $0.27<\bar{q}_6<0.31$ (blue) and $0.27<\bar{q}_6<0.34$ (magenta) at I, II and III. The vertical dashed lines separate FCC/HCP and BCC-like local environments. The gradual shift of the distribution towards FCC/HCP-like local structures ($w_6<0$) on increasing $\bar{q}_6$ is suppressed as one moves from I to III. (bottom) The conditional probability distributions of order parameter $\bar{w}_4$ for fluid particles with $\bar{w}_6\le 0$ and for the same $\bar{q}_6$ ranges as in the top panel.}
\label{fig_3s}
\end{figure}

In Fig.~\ref{fig_3s} (top panel), we show the evolution of the local $\bar{w}_6$ distribution with the extent of crystallinity which is quantified using the local $\bar{q}_6$ order parameter in the metastable fluid phase. At I, the system becomes increasingly enriched with FCC/HCP-like environments ($\bar{w}_6<0$) on including the particles with the higher $\bar{q}_6$ values. A similar behavior is observed at II, which is consistent with the recent observation for the Gaussian Core Model system~\cite{tanaka_gcm_2012}. In contrary, at III, the population is slightly biased towards BCC-like environments ($\bar{w}_6>0$). To further analyze the polymorphic identity of the particles with $\bar{w}_6 < 0$, in the bottom panel of Fig.~\ref{fig_3s}, we present the distribution of local $\bar{w}_4$ for the particles with $\bar{w}_6 < 0$ on varying the range of $\bar{q}_6$ (same as in the top panel). This order parameter distinguishes the FCC-like ($\bar{w}_4 < 0$) and HCP-like ($\bar{w}_4 > 0$) local environments. As the figure suggests, at all the three conditions, the distributions are biased towards FCC-like environments ($\bar{w}_4 < 0$). That is, the particles with $\bar{w}_6<0$ are dominated by the FCC-like local environments ($\bar{w}_4 < 0$), as is also evident from Figs.~\ref{fig_2},~\ref{fig_3} and~\ref{fig_4}(b).

The results shown in Fig.~\ref{fig_4} and Fig.~\ref{fig_3s} highlight the underlying connection between the free energy cascade and the nature of the fluctuations in the metastable fluid and confirm recent observations that the information about the polymorph selection is indeed encoded in thermal fluctuations of local bond-orientational order parameters in the metastable fluid phase.

\begin{figure}[!ht]
\includegraphics[width=\linewidth]{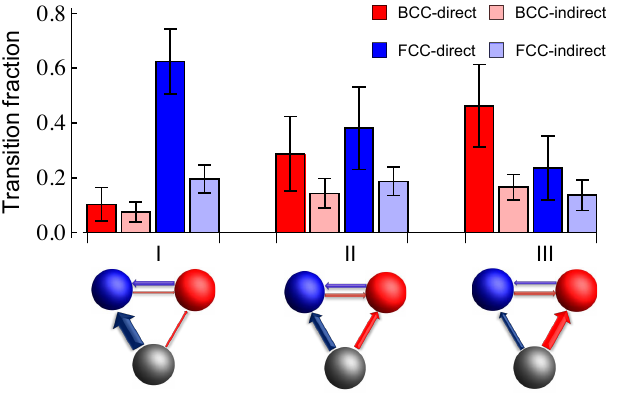}
\caption{The fraction of solid-like particles in the critical cluster formed via direct and indirect routes at state points I, II and III in the phase diagram. BCC-direct (FCC-direct) denotes the BCC-like (FCC-like) particles formed directly from their fluid-like environment and BCC(FCC)-indirect denotes BCC(FCC)-like particles formed through an indirect solid-solid transition mediated route --- fluid$\rightarrow$FCC$\rightarrow$BCC (fluid$\rightarrow$BCC$\rightarrow$FCC). Note that at I, FCC-direct path is the dominant contributor, and at III, BCC-direct path is the dominant contributor to the formation of solid-like particles in the critical cluster. At II, we observe competitive direct appearance of FCC and BCC-like particles from respective fluid-like environments. The transition fractions were obtained by averaging over $100$ independent trajectories. The lower panel depicts a schematic representation where grey, blue and red spheres indicate fluid, FCC and BCC-like particles, respectively, and the width of the arrow for each step is proportional to its weight.}
\label{fig_5}
\end{figure}

\subsection{How does the composition of critical cluster emerge via spontaneous fluctuations in the metastable fluid?}
Finally, as the composition of the clusters controls the free energy cost of their formation (Fig.~\ref{fig_2} and Fig.~\ref{fig_3}), we address the question of the selection and emergence of the critical cluster composition (fraction of different solid-like particles in the critical cluster) by fluctuations off the metastable fluid phase. Following the history of each solid-like particle in the critical cluster one can gain a mechanistic understanding of how each of the solid-like particles in the critical cluster eventually form from fluid-like environment. The transformation of a fluid-like particle to the solid-like can occur either through (a) direct transformation of the fluid to the final solid-like environment or (b) indirect pathway where fluid-like particle first transforms to BCC(FCC) and then to FCC(BCC). We computed the weights of these two pathways for each of the solid-like particle belonging to the critical cluster (Fig.~\ref{fig_5}). At I, the majority of the solid-like particles in the critical cluster ($\sim 60\%$) forms via the direct fluid$\rightarrow$FCC pathway and only 20\% via the indirect (fluid$\rightarrow$BCC$\rightarrow$FCC) pathway (Fig.~\ref{fig_5}, top panel). On the other hand, at III, the majority of transitions ($\sim50\%$) is via the direct fluid$\rightarrow$BCC pathway. At II, however, we observe a competitive direct appearance of FCC and BCC-like particles (ca. 30\% and 40\%, respectively). In all the three cases, the solid-like particles formed via indirect pathways are only 15-20\%. This observation suggests that the composition of the critical cluster is predominantly guided by the direct transformation of the fluid-like particles to the solid-like rather than indirect pathways mediated by solid-solid transformations. The lower panel of Fig.~\ref{fig_5} depicts a schematic representation of the whole process.  

\section{Conclusions} \label{conclusions}
Using hard-core repulsive Yukawa as a model charge-stabilized colloidal system, we demonstrate here the kinetic origin of preferential formation of the BCC phase even though FCC is  thermodynamically the most stable phase, thus unambiguously justifying the age-old observation of Alexander and McTague~\cite{mctag}. In this process, this study brings out the true essence of the Ostwald's step rule~\cite{ostwald_1,ostwald_2}. We further show that the nature of the local bond-orientational oder parameter fluctuations in the metastable fluid phase as well as the composition and size of the critical nucleus depend delicately on the relative stability of the intermediate BCC phase. The composition of the critical cluster is guided by the direct transformation of fluid-like particles to the solid-like rather than indirect pathways involving solid-solid transformations. In addition, the results obtained in this work qualitatively validate the predictions of our recent phenomenological classical DFT~\cite{bagchi_osr_2013}. As this theory is not specific to any particular system or inter-particle interaction potential, we anticipate that, irrespective of the system, if the nature of the free energy landscape is like the one considered here, the results obtained in this work should hold true. We also anticipate that this study could provide important insights into the synthesis of polymorphs of desired structures and properties by controlled change of the free energy landscape either through changing the thermodynamic conditions or altering the inter-particle interactions (especially in macro- and mesoscale systems)~\cite{whitelam_2012, sciortino_natphys_2014, frenkel_pnas_2014, frenkel_prl_2014}. 

\begin{acknowledgments}
B.B. gratefully acknowledges support from the Department of Science and Technology (DST,
India) and Sir J. C. Bose fellowship for providing partial financial
support. 
\end{acknowledgments}

\bibliography{jcs}

\end{document}